\documentclass[journal,lettersize]{IEEEtran}

\usepackage{graphicx} 
\usepackage{xcolor}
\usepackage{amsmath} 
\usepackage{amsfonts}
\usepackage{cite}
\usepackage{array}
\usepackage{xcolor}
\usepackage{tikz}
\usetikzlibrary{spy,arrows}
\usepackage{pgfplots}
\usepackage{tkz-euclide,subfigure}
\usepackage{epsfig,amssymb}
\usepackage{svg}
\pgfplotsset{compat=1.18}

\usepackage[acronym]{glossaries}
\glsdisablehyper
\newacronym{ai}{AI}{artificial intelligence}
\newacronym{dt}{DT}{digital twin}
\newacronym{ppi}{PPI}{prediction-powered inference}
\newacronym{marl}{MARL}{multi-agent reinforcement learning}
\newacronym{aod}{AoD}{angle of departure}
\newacronym{los}{LoS}{line-of-sight}
\newacronym{nlos}{NLoS}{non-line-of-sight}
\newacronym{bs}{BS}{base station}
\newacronym{erm}{ERM}{empirical risk minimization}
\newacronym{p-erm}{P-ERM}{pseudo-empirical risk minimization}
\newacronym{dr}{DR}{doubly-robust}
\newacronym{cdr}{CDR}{context-aware doubly-robust}
\newacronym{inr}{INR}{implicit neural representation}
\newacronym{cppi}{CPPI}{context-aware PPI}
\newacronym{ckm}{CKM}{channel knowledge map}
\newacronym{em}{EM}{expectation-maximization}
\newacronym{nf}{NF}{network function}

\begin{document}
\bstctlcite{IEEEexample:BSTcontrol} 

\title{How to Bridge the Sim-to-Real Gap in Digital Twin-Aided Telecommunication Networks}

\author{Clement~Ruah, Houssem~Sifaou, Osvaldo~Simeone, and Bashir~M.~Al-Hashimi
\vspace{-0.5cm} 
\thanks{C. Ruah, H. Sifaou and O. Simeone are with the Institute for Intelligent Networked Systems (INSI) at Northeastern Unversity London, London E1 8PH, UK (email: \{c.ruah, h.sifaou, o.simeone\}@northeastern.edu).
B. M. Al-Hashimi is with the Department of Engineering,  King’s College London, London WC2R 2LS, UK. 
This work was supported by the European Research Council (ERC) under the European Union’s Horizon Europe Programme (101198347), the European Union’s Horizon Europe Project CENTRIC (101096379),  by~an Open Fellowship of the EPSRC (EP/W024101/1), and~by the EPSRC project (EP/X011852/1).}
}

\maketitle

\begin{abstract}
Training effective artificial intelligence models for telecommunications is challenging due to the scarcity of deployment-specific data. Real data collection is expensive, and available datasets often fail to capture the unique operational conditions and contextual variability of the network environment. Digital twinning provides a potential solution to this problem, as simulators tailored to the current network deployment can generate site-specific data to augment the available training datasets. However, there is a need to develop solutions to bridge the inherent simulation-to-reality (sim-to-real) gap between synthetic and real-world data. This paper reviews recent advances on two complementary strategies: 1) the calibration of \glspl{dt} through real-world measurements, and 2) the use of sim-to-real gap-aware training strategies to robustly handle residual discrepancies between digital twin-generated and real data. For the latter, we evaluate two conceptually distinct methods that model the sim-to-real gap either at the level of the environment via Bayesian learning or at the level of the training loss via prediction-powered inference.
\glsreset{dt}
\end{abstract}

\section{Introduction}\label{sec:intro}

Driven by the continued growth of computing resources and training datasets, \gls{ai} research is widely considered to be in the scaling era, which is focused on the development of general-purpose models that exhibit emergent capabilities. While this trend has yielded impressive results for many tasks, particularly in the domain of language modeling, it poses unique challenges when applied to engineering domains such as telecommunication networks.

One major concern is the energy cost: training and deploying large-scale \gls{ai} models in data centers consumes significant amounts of power, raising sustainability and operational efficiency issues. Another critical challenge is the highly {contextual} nature of telecommunication data. The performance of wireless networks is highly sensitive to specific deployment factors, including the propagation environment, network topology, and traffic patterns. These factors vary widely across different settings, making it impractical to collect sufficient real-world data to train robust \gls{ai} models for every possible context.

A promising approach to address data scarcity in telecommunications is the use of {\glspl{dt}}---high-fidelity simulators or emulators of network environments that can efficiently generate synthetic data tailored to specific deployments. By training \gls{ai} models on a combination of real and synthetic data, it becomes possible to improve generalization performance compared to training solely on the limited real-world data typically available for a given deployment.
Examples of \glspl{dt} include  NVIDIA's Sionna library, which offers an efficient simulator of the propagation environment via ray tracing \cite{hoydis2024learning}, with recent extensions also incorporating networking aspects \cite{pegurri2024toward}. \Glspl{dt} may also encompass real-world elements such as software-defined radios \cite{polese2024colosseum}. 

 \begin{figure*}
    \centering
    \includegraphics[keepaspectratio=true, width=0.85\textwidth]{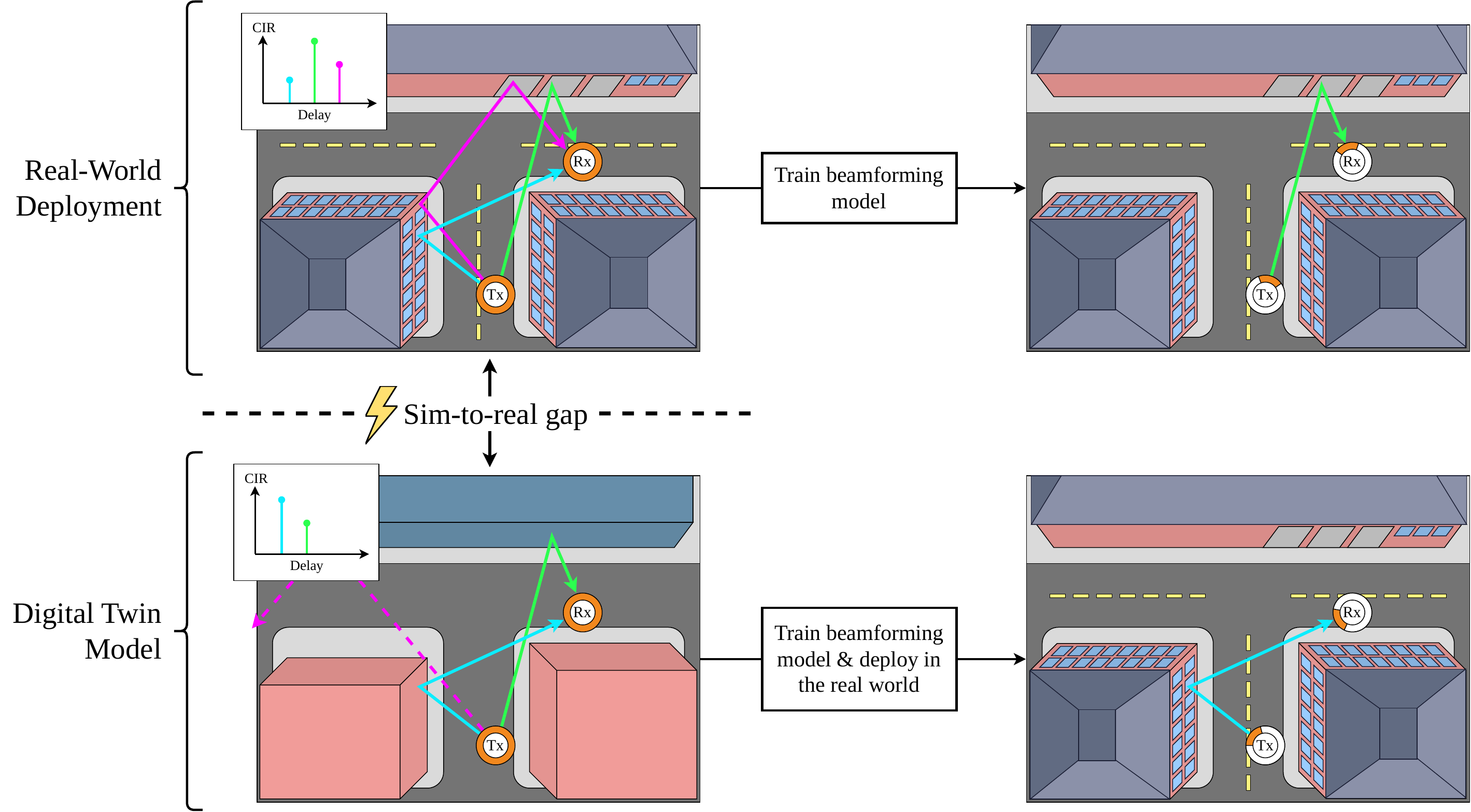}
    \caption{(left) Illustration of the sim-to-real gap between (top) a real-world urban wireless deployment and (bottom) a ray tracing-based digital twin (DT). The DT model may yield incorrect channel impulse responses (CIRs) due to a mismatch between the real geometry and material parameters and the ones assumed by the simulator. (right) When using the DT-predicted CIRs to train a beamforming model, the sim-to-real gap may cause the selection of a sub-optimal beam.}
    \label{fig:overview}
\end{figure*}

However, a key challenge when training on a mix of real and synthetic data is that the \gls{dt} may not perfectly reflect the real data distribution. There is typically a non-negligible {simulation-to-reality (sim-to-real) gap} due to modeling simplifications or uncertain environment parameters. 
The left part of Fig.~\ref{fig:overview} illustrates this gap for a ray tracing \gls{dt} engine (bottom) that fails to correctly simulate the true wireless propagation conditions of the real urban environment (top) due to its simplified virtual geometry and its lack of knowledge of the real electromagnetic parameters of the objects in the scene.
If an \gls{ai} model is trained naively on synthetic data produced by such a \gls{dt}, these discrepancies can cause degraded performance when the model is deployed into the real-world system. In the right part of the example in Fig.~\ref{fig:overview}, a beamforming model trained on synthetic data (bottom) may fail to select the best beam (top).

\begin{figure*}
    \centering
    \subfigure[Digital twin calibration]{
        \includegraphics[keepaspectratio=true, width=0.85\textwidth]{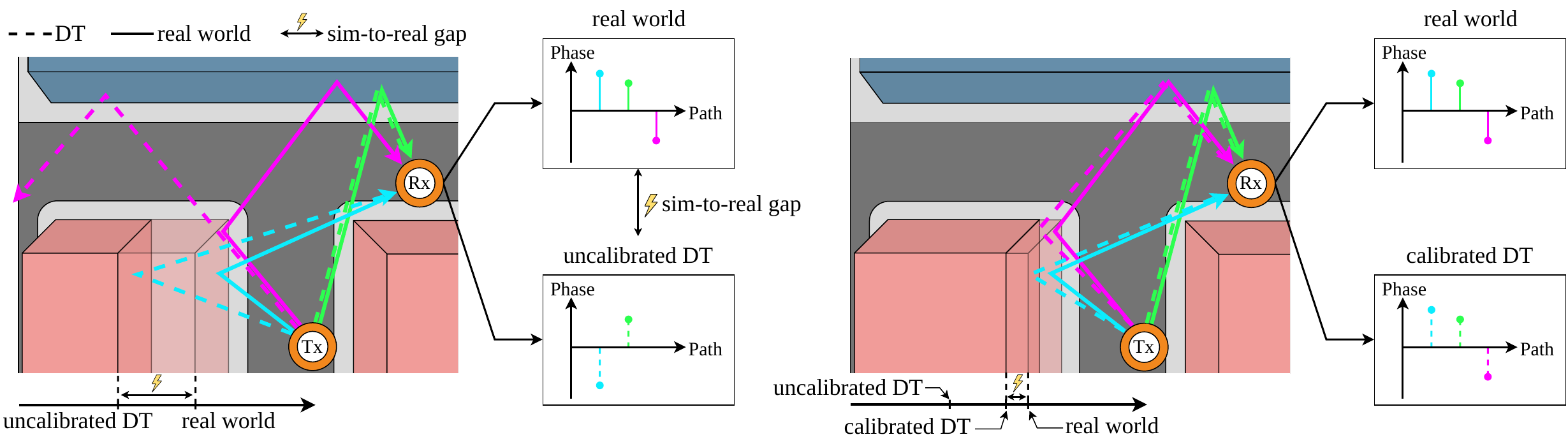}
        \label{subfig:method_calibration}
    }
    \hrule
    \subfigure[Modeling the sim-to-real gap on the environment]{
        \includegraphics[keepaspectratio=true, width=0.34\textwidth]{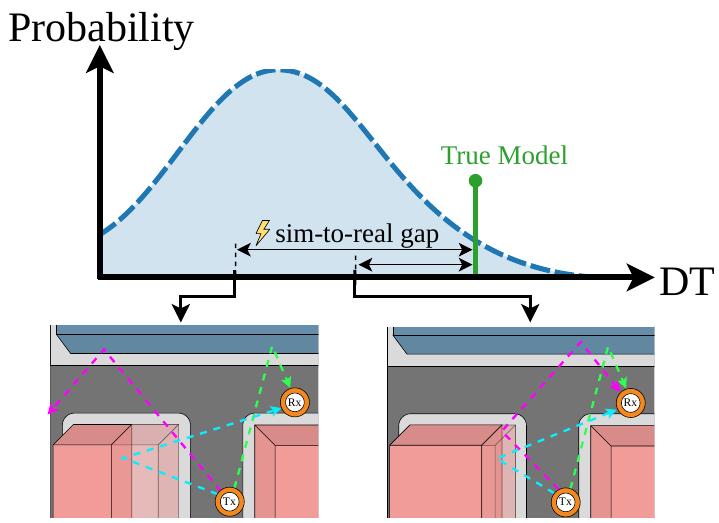}
        \label{subfig:method_bayesian}
    }
    \hfill
    \vrule
    \hfill
    \subfigure[Modeling the sim-to-real gap on the training objective]{
        \includegraphics[keepaspectratio=true, width=0.54\textwidth]{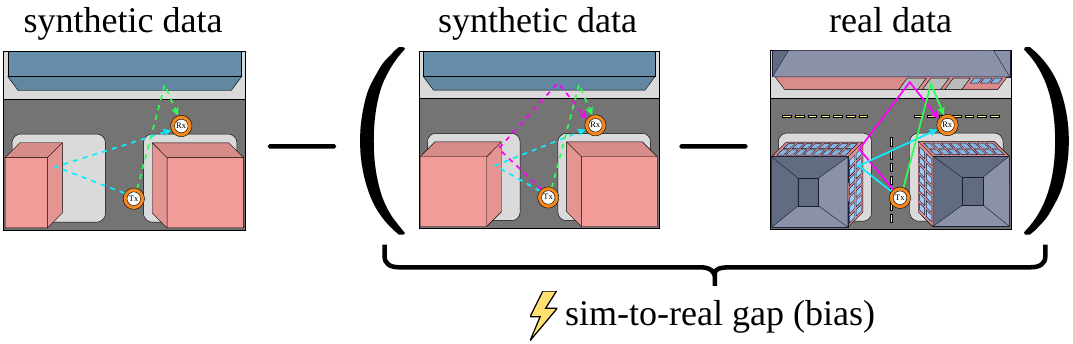}
        \label{subfig:method_ppi}
    }
    \caption{
    As detailed in this paper, the sim-to-real gap can be partially bridged via three main complementary strategies: (a) Digital twin calibration (Sec.~\ref{sec:calibrate_dt}), which uses real-world measurements to directly improve the adherence of the DT to the real world; (b)-(c) Robust AI training, which accounts for the residual sim-to-real gap by modeling the uncertainty on the true environment via Bayesian inference (panel (b), Sec.~\ref{sec:bayesian_dt}), or by correcting the bias caused by the sim-to-real gap on the training objective (panel (c), Sec.~\ref{sec:ppi_dt}). (Dashed lines indicate DT-generated synthetic data, solid lines represent real-world measurements.)
    }
    \label{fig:sim_to_real_methods}
\end{figure*}

How can we bridge the sim-to-real gap to ensure that AI models trained with \glspl{dt} remain reliable in real deployments? 
As illustrated in Fig.~\ref{fig:sim_to_real_methods}, this article explores three complementary strategies:

\noindent $\bullet$ \textbf{Calibration of \glspl{dt}} (Sec.~\ref{sec:calibrate_dt}): We first discuss techniques for calibrating \glspl{dt} using real-world measurements, with the goal of directly improving simulator fidelity and reducing bias in synthetic data generation. For example, in the setting illustrated in Fig.~\ref{fig:overview}, calibration methods may improve the description of the scene geometry and/or the estimate of the material parameters in the scene by matching real-world observations with synthetic data (see Fig.~\ref{subfig:method_calibration}).

\noindent $\bullet$ \textbf{Robust AI training via \glspl{dt}} (Sec.~\ref{sec:bayesian_dt} and \ref{sec:ppi_dt}): Calibration methods may reduce the sim-to-real gap, but they cannot fully remove it owing to the inherent bias entailed by the use of limited-complexity models and simulators. Therefore, even when using a calibrated \gls{dt}, it is important to design training methods that explicitly account for the residual sim-to-real gap in the synthetic data generated by the \gls{dt}. These techniques fall into two distinct, and complementary, categories:
\begin{itemize}
    \item[$\circ$] \textbf{Modeling the sim-to-real gap on the environment} (Sec.~\ref{sec:bayesian_dt}): A first approach is to explicitly model the sim-to-real gap directly at the level of the \gls{dt} model. Accordingly, the underlying \gls{dt} model is treated as a random quantity, on which the \gls{dt} refines its prior knowledge by using the available data. In the ray tracing example, the geometry and/or the electromagnetic properties of the objects in the scene may be modeled as random variables, whose uncertainty describes the residual ambiguity about the mismatch between real and simulated environments (see Fig.~\ref{subfig:method_bayesian}).  This probabilistic modeling, formalized by Bayesian inference,  makes it possible to generate synthetic data from multiple plausible \gls{dt} models, thus accounting for the sim-to-real gap resulting from prior knowledge and from limited data.
    \item[$\circ$] \textbf{Modeling the sim-to-real gap on the training objective}  (Sec.~\ref{sec:ppi_dt}): Synthetic data produced by a \gls{dt} are typically used to train \gls{ai} models, e.g., the beamforming mapping in Fig.~\ref{fig:overview}. In these settings, training is often done by minimizing a training loss evaluated using both real and synthetic data. In light of this, a complementary approach to devising reliable \gls{ai} training strategies in the presence of \glspl{dt} is to mitigate the sim-to-real gap directly at the level of the training loss. These techniques seek to reduce the \gls{dt} bias under sim‑to‑real gaps by incorporating available real‑world observations. A prominent approach to achieve this is {\gls{ppi}} \cite{angelopoulos2023prediction} (see Fig.~\ref{subfig:method_ppi}).
\end{itemize}

The three solutions described above are orthogonal and complementary.
Accordingly, they can be applied on the same system in order to reap their distinct benefits.
For instance, calibrating the \gls{dt} reduces the need for mitigating the sim-to-real gap; while accurately modeling the sim-to-real gap on the model makes it less critical to address the sim-to-real gap on the training objective.
Given that all methods are based on different principles, they are described and evaluated separately in this article.
The benefits, data requirements, computational costs, and reusability across \glspl{nf} of the methods are summarized in Table~\ref{tab:methods_summary}.

As a final remark, we emphasize that this paper focuses on leveraging \glspl{dt} for data augmentation in AI model training. While this represents an important use case, as highlighted by O-RAN \cite{oran2024dtran}, \glspl{dt} can also enable real-time operational tasks, including, for instance, channel and link quality prediction \cite{DTrev}.
Such real-time applications require continuous synchronization mechanisms to manage the real-to-sim gap at fine temporal granularities \cite{hashash2022edge}. In contrast, our work addresses \gls{dt} calibration at a non-real-time scale, focusing on training AI models using \glspl{dt} that capture long-term environmental statistics and operating conditions.

\begin{table}[htbp]
  \caption{Comparison of sim-to-real gap reduction methods}
  \label{tab:methods_summary}
  \centering
  \renewcommand{\arraystretch}{1.3}
  \setlength{\extrarowheight}{1pt}
  \setlength{\tabcolsep}{5pt}
  \begin{tabular}{p{0.14\columnwidth} p{0.13\columnwidth} p{0.18\columnwidth} p{0.12\columnwidth} p{0.22\columnwidth}}
    \hline
        \centering\textbf{Method} &
        \centering\textbf{Benefits} &
        \centering\textbf{Data requirement} & 
        \centering\textbf{Compute cost} &
        \centering\arraybackslash\textbf{Deployment type} \\
    \hline
        Calibration &
        Improved DT &
        Large dataset from specific deployment &
        Training &
        Deployment-specific, reusable across NFs \\
    \hline
        Modeling environment gap &
        Uncertainty quantification &
        Limited data from multiple deployments &
        Training + sampling &
        Deployment-specific, reusable across NFs \\
    \hline
        Modeling train  ob-jective gap &
        Debiased training objective &
        Limited data for specific NF &
        Training &
        NF-specific \\
    \hline
  \end{tabular}
\end{table}

\section{Calibrating Digital Twins} \label{sec:calibrate_dt}

Emerging physics-based \glspl{dt} of telecommunication networks often include models of the wireless propagation environment using {ray tracing} \cite{hoydis2024learning}. 
In a ray tracing simulator, the environment is defined by a three-dimensional {geometry} along with a vector of {material parameters} describing the reflection, refraction, and diffraction characteristics of surfaces in the scene.
The simulator models the propagation of radio waves by tracing the paths connecting transmitters to receivers, keeping track of the interactions of each path with the objects in the environment.  
The multi-path parameters returned by the simulator can then be leveraged to produce a simulated channel response or to estimate the received power at given locations (see Fig.~\ref{fig:overview}).

In practice, however, a sim-to-real gap is inevitable: the specified geometry of the scene might miss objects or report inaccurate placements or dimensions, while material parameters may be unknown or wrongly specified. These mismatches lead to biases in the synthetic data generated by the \gls{dt}.
Therefore, as discussed in this section, it is important to reduce the sim-to-real gap caused by model misspecification by leveraging available real-world data to calibrate the \gls{dt}.

While this section focuses on the calibration of physics-based models, data-driven models, such as dynamic systems, can also benefit from periodic calibration based on current data. This can be done using standard continual learning tools, which are not covered here \cite{hashash2022edge}.

\subsection{Material and Parameter Calibration via Differentiable Simulation}\label{sec:mat}

One way to calibrate a \gls{dt} is to {learn the unknown material parameters} by comparing simulation outputs to real measurements. To elaborate, we denote as $\rho$ the material parameters for the objects present in the scene. These include permittivities, conductivities, and scattering patterns. In practice, it is also possible to define an \gls{inr} that associates each position in the scene with a vector $\rho$ \cite{hoydis2024learning}.

To support calibration, the network operator deploys a few test transmitters and receivers in the real network, and records the channel measurements $H$. Given a vector of model parameters $\rho$, one can then run the ray tracing simulator to yield a prediction $\hat{H}(\rho)$ of the channel measurements. Using this information, calibration can be formulated as an inverse problem: find the parameters $\hat{\rho}$ that cause the simulated measurement $\hat{H}(\hat{\rho})$ to best match the real observations $H$. This criterion is averaged over the available data set in order to obtain statistically sound estimates of the parameters $\rho$.

Recent research has shown that this inverse problem can be efficiently addressed by leveraging  {differentiable} ray tracing software, and applying gradient-based optimization techniques to adjust the material parameters \cite{hoydis2024learning}. This approach treats the \gls{dt} as a trainable model mapping material parameters to multi-path parameters, making it possible to differentiate through the ray tracing engine. This, in turn, supports the use of gradient descent to infer material properties $\rho$ that minimize the error between simulated and measured channel responses. 

\subsection{Phase Error-Aware Calibration}

Calibrating material parameters is useful, but it cannot fully resolve biases caused by a misspecified geometry of the objects in the scene.
For instance, as seen in the blue path of Fig.~\ref{subfig:method_calibration}, minor errors in building shapes or additional objects not present in the \gls{dt} can cause significant {phase errors} in the simulated propagation paths.
In fact, even a small displacement of an object---of the order of a fraction of the wavelength---can cause the phase of a reflected path to shift, leading to constructive or destructive interference patterns that do not align with reality \cite{ruah2024calibrating}.
Owing to these unaccounted phase errors, traditional calibration methods that only tweak material values $\rho$ may fail to produce an accurate simulator.

To tackle this issue, a promising approach is to devise calibration schemes that are aware of the presence of phase errors caused by a sim-to-real gap in the geometry of the environment.
The core idea behind {phase error-aware calibration}, as proposed in \cite{ruah2024calibrating}, is to iteratively estimate and correct phase errors in the simulated channel.
In this method, the \gls{dt}'s prediction is augmented with per-path phase error terms that account for unmodeled geometry nuances.
The errors are illustrated by the phase difference of the blue path in Fig.~\ref{subfig:method_calibration} between the ``real world'' (solid line) measurement and the ``uncalibrated \gls{dt}'' (dashed line) prediction.

The phase error estimates are used to correct the phase of the predicted paths to match real world conditions, as shown in the ``calibrated \gls{dt}'' phase panel of Fig.~\ref{subfig:method_calibration}.
The resulting calibration algorithm alternates between phase error correction and material calibration within an \gls{em} framework: in the E-step, the algorithm infers the likely phase errors for each propagation path by comparing the simulation with real measurements; while the M-step updates the simulator parameters $\rho$ to better fit the observed data, effectively absorbing the phase errors into the model.

\begin{figure}
    \centering
    \includegraphics[keepaspectratio=true, width=0.85\columnwidth]{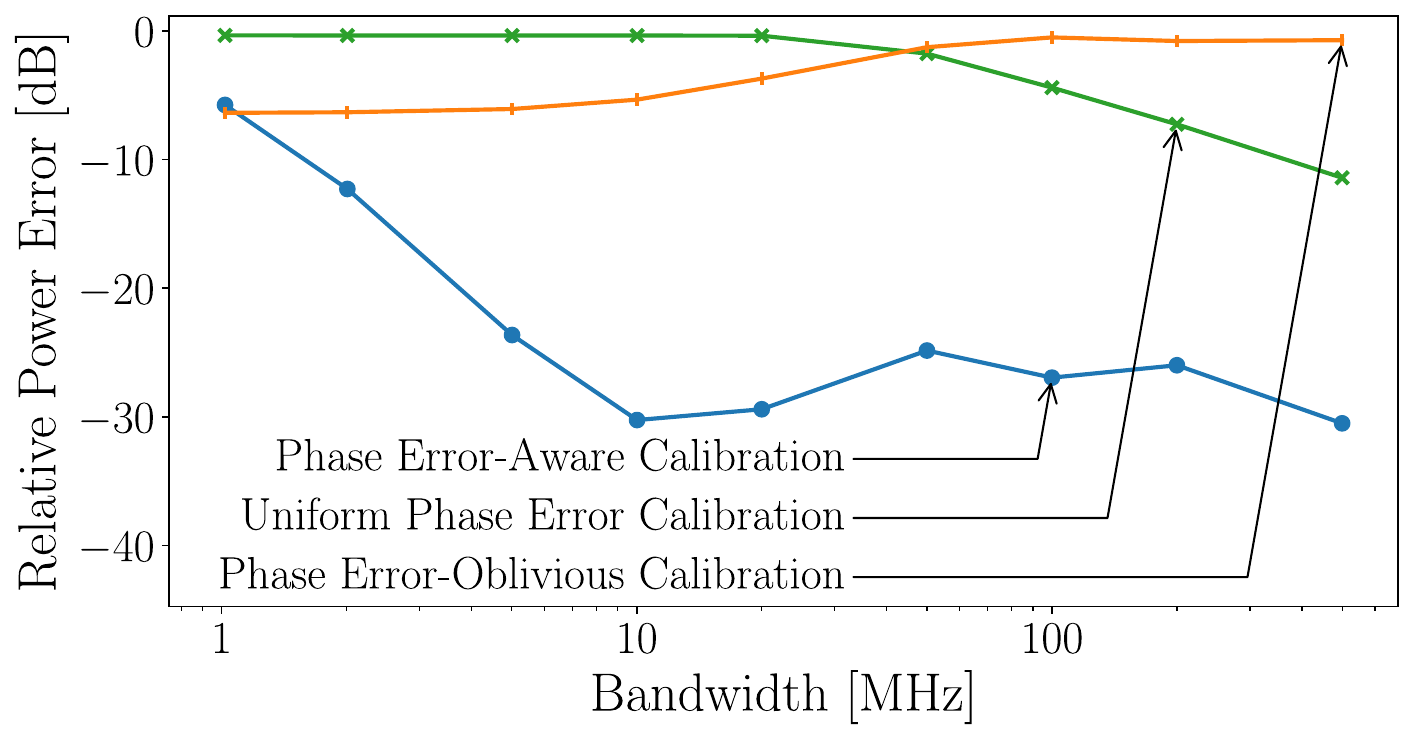}
    \caption{
    Relative power prediction errors of ray tracing-based DTs calibrated using phase error-oblivious (orange), uniform phase error (green), and phase error-aware (blue) calibration methods, where calibration is conducted using channel measurements at different bandwidths (figure adapted from \cite{ruah2024calibrating}).
    }
    \label{fig:calibration_error_vs_bandwidth}
\end{figure}

Fig.~\ref{fig:calibration_error_vs_bandwidth} shows the performance of different calibration techniques for a canyon scenario with \gls{los} blockage, resulting in two propagation paths connecting the transmitter to the receiver by reflecting on each of the canyon's walls. In this setting, geometry mismatch at the \gls{dt} is induced by a misplacement of one of the walls with respect to the true scene \cite{ruah2024calibrating}.
The calibration procedures are conducted under multiple levels of available bandwidth for channel measurements, and the results quantify the relative error between the measured power and the power predicted by the calibrated \gls{dt}.
We compare the described phase error-aware approach with the following alternative methods: (\emph{i}) \emph{phase error-oblivious calibration}, which disregards phase errors due to geometry mismatch when matching simulated and measured channel responses; and (\emph{ii}) \emph{uniform phase error calibration}, which assumes the phases of the predicted paths to be uniformly distributed when matching delay-angle projections of the predicted and measured channel powers during calibration.
All methods tune the material parameters $\rho$ via gradient-based optimization \cite{hoydis2024learning} (see Sec. \ref{sec:mat}).

By disregarding phase errors, phase-oblivious calibration is seen to be unable to provide satisfactory results across all bandwidth levels. Uniform phase error calibration yields improved results as the bandwidth levels approach the delay-resolution needed to separate the contributions of the interfering paths, allowing for material parameters calibration by matching the energy of each path individually.
Finally, phase error-aware calibration can dramatically reduce the power prediction error as compared to the two baseline methods, even at lower bandwidth levels. This demonstrates that explicitly accounting for local phase error during calibration can significantly improve the reliability of the ray tracer's estimates. This advantage is especially pronounced in scenarios with a rich multipath structure, in which small geometry differences would otherwise lead to large interference errors.

\begin{figure*}
    \centering
    \includegraphics[keepaspectratio=true, width=0.85\textwidth]{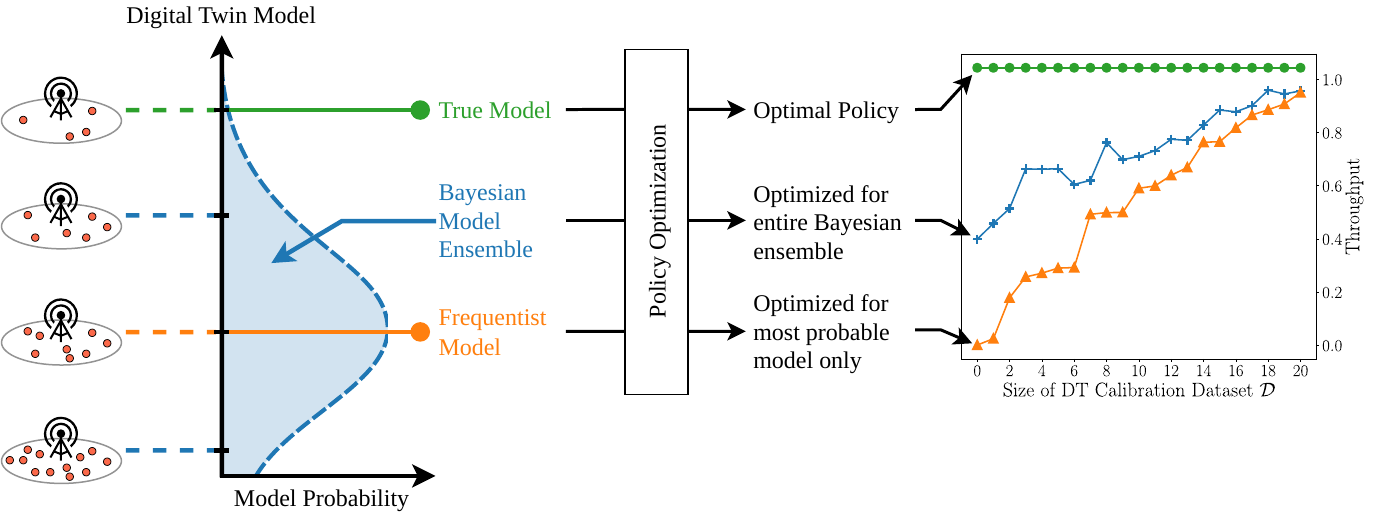}
    \caption{
    Training a multiple access protocol using synthetic data from a digital twin (DT) modeling the access channel: (left) a Bayesian formulation of DT learning assigns different configurations (interference levels) of the DT to a probability distribution that depends on prior knowledge and available data.  In contrast, a frequentist approach would only select the highest-probability model. (center) While the frequentist DT trains the policy inside the single selected model, the Bayesian DT generates data from multiple models, weighting the importance of multiple plausible DT models during policy optimization. (right) When calibration data is in limited supply, explicitly modeling the sim-to-real gap via uncertainty-aware Bayesian methods can significantly improve the performance of models trained using synthetic data from the DT (adapted from \cite{ruah2023bayesian}).  
    }
    \label{fig:bayesian_dt}
\end{figure*}

\section{Modeling the Sim-to-Real Gap on the Environment} \label{sec:bayesian_dt}

Through calibration, the synthetic data produced by the \gls{dt} becomes a much closer proxy for real data. Calibration thus attacks the sim-to-real gap at its source, reducing the bias in any model trained on the \gls{dt}. Nonetheless, in practice, there will always remain a residual uncertainty or bias, contributing to a non-negligible sim-to-real gap. For example, a ray tracing simulator will inevitably leverage simplified geometric and electromagnetic configurations. For this reason, it is also important to design techniques that make the \gls{ai} training process itself robust to any residual sim-to-real gap.

Rather than further refining the simulator, which may not be feasible, the goal of the next two sections is to robustly train \gls{ai} models using synthetic data generated by a \gls{dt}. A principled way to account for the sim-to-real gap during training is to explicitly model uncertainty about the true environment at the \gls{dt} itself before generating synthetic data. This way, rather than assuming the digital replica of the environment provided by the \gls{dt} to be ideal, we treat the parameters that define the \gls{dt}—such as the material or geometric parameters discussed in Sec.~\ref{sec:calibrate_dt}—as random variables within a Bayesian perspective.

In this Bayesian framework, as illustrated in Fig. \ref{subfig:method_bayesian}, one maintains a probability distribution over the possible configurations of the \gls{dt}, reflecting the epistemic uncertainty on the environment under the limited available data. Initially, this uncertainty is encoded in a prior distribution, which expresses our knowledge---or lack thereof---before observing any real-world data. As we collect real measurements, the distribution is updated to a posterior distribution, thereby refining the \gls{dt}'s understanding of the environment and modeling any residual sim-to-real gap. Instead of relying on a single ``best-guess'' \gls{dt} model for synthetic data generation, with Bayesian learning one can generate synthetic data from an ensemble of \gls{dt} models sampled from the posterior distribution, enabling a more robust learning of \gls{ai} policies. By sampling models from the posterior distribution, during data generation, the weight given to each \gls{dt} model depends on the belief assigned by the system to the given model based on the available information.

As an exemplifying use case, following \cite{ruah2023bayesian}, we now focus on a system in which multiple devices coordinate to access a shared wireless medium using a protocol learned via \gls{marl}. The \gls{marl} protocol is trained using synthetic data from a \gls{dt}. If we were to use a single  \gls{dt} model to simulate the environment, as done in a conventional frequentist implementation, the resulting policy might perform well under that specific model, but it could fail to generalize to the real-world system due to the real-to-sim gap. In contrast, the Bayesian approach samples multiple plausible environments from the posterior distribution, training the \gls{marl} policy across a number of different \gls{dt} models. This leads to a policy that is robust under the sim-to-real gap modeled at the \gls{dt}, achieving better generalization to the true environment.

Fig.~\ref{fig:bayesian_dt} elaborates on the operation and on the advantages of the Bayesian methodology for a simulated multiple access setting composed of four devices that communicate with a single \gls{bs} under an unknown shared channel. The shared channel attributes lower probabilities of success to the delivery of packets when multiple devices transmit within the same time slot \cite{ruah2023bayesian}. Accordingly, the left panel of Fig.~\ref{fig:bayesian_dt} illustrates a posterior distribution over different \gls{dt} models that assume varying levels of interference, resulting in distinct likelihoods to the number of received packets given the number of concurrently transmitted packets. The central panel contrasts the training processes of different methodologies. The conventional frequentist learning approach relies on a single maximum a posteriori model, while the Bayesian learning framework integrates across a range of likely models, weighting each model according to its posterior probability. As shown in the right panel, this yields superior performance for the policy trained using synthetic data, especially in data-scarce regimes where the model uncertainty about the environment dynamics is high.

Bayesian \gls{dt} training can be implemented through various practical methods. Examples include Monte Carlo dropout, deep ensembles, and variational inference techniques that approximate the posterior distribution \cite[Ch.~12]{simeone2022machine}.
The actual benefits of Bayesian learning hinge on reasonable choices for prior distribution and for the environment model, although there is a vast literature on methods that can ensure robustness to model misspecification \cite{zecchin2023robust}.
Another limitation of Bayesian learning is the increased computational complexity: sampling multiple environments, training across a population of models, and maintaining distributions over parameters may introduce significant overhead.
Nonetheless, in many critical applications where reliability and safety are paramount, this additional cost often offers a worthwhile trade-off.

\begin{figure*}
    \subfigure[Cross-PPI]{
        \centering
        \includegraphics[keepaspectratio=true, width=0.68\linewidth]{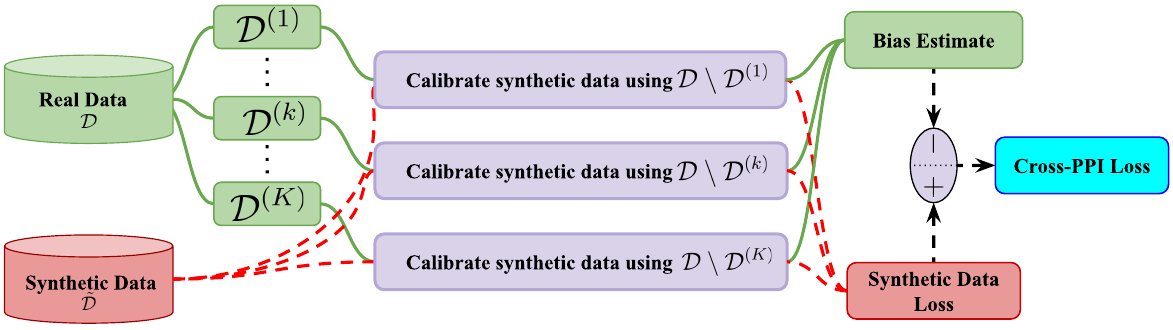}
    }
    \subfigure[Beam alignment via CKM]{
    \centering
        \begin{tikzpicture}[scale=0.54]
            \begin{axis}[
                tick align=outside,
                tick pos=left,
                xlabel style={font=\large},
                ylabel style={font=\large},
                tick label style={font=\large},
                x grid style={white!69.0196078431373!black},
                xlabel={Proportion of real data},
                xmajorgrids,
                xmin= 512/38038, xmax=4100/38038,
                xtick style={color=black},
                y grid style={white!69.0196078431373!black},
                ylabel={Channel capacity (bps/Hz)},
                ymajorgrids,
                ymin=3.5, ymax=5.5,
                ytick style={color=black},
                grid=major,
                scaled ticks = true,
                legend pos = south east,
                legend style={nodes={scale=0.8, transform shape}},
                grid style=densely dashed,
            ]

            \addplot[
                thick, red, mark=star, line width=2pt
            ] coordinates {
                (128/38038, 2.514505359522904)
                (256/38038, 2.961611773529421)
                (512/38038, 3.4743600451567844)
                (1024/38038, 3.7288443195443928)
                (2048/38038, 4.003198497178796)
                (4096/38038, 4.443089397969243)
            };
            \addlegendentry{ERM}

            \addplot[
                thick, black, dashed, mark=none, line width=2pt
            ] coordinates {
                ({128/38038}, 1.939229513428491)
                ({256/38038}, 2.218801517818342)
                ({512/38038}, 2.825604075466892)
                ({1024/38038}, 3.6582947308801166)
                ({2048/38038}, 4.047121394096558)
                ({4096/38038}, 4.409081553728858)
            };
            \addlegendentry{P-ERM}

            \addplot[
                thick, violet, mark=triangle, mark size=2, line width=2pt
            ] coordinates {
                ({128/38038}, 2.8718675582774065)
                ({256/38038}, 3.3895610809050414)
                ({512/38038}, 3.6760500342169613)
                ({1024/38038}, 4.4109511203354606)
                ({2048/38038}, 4.6578578646712865)
                ({4096/38038}, 4.840696840513544)
            };
            \addlegendentry{PPI}

            \addplot[
                thick, blue, mark=square, mark size=2, line width=2pt
            ] coordinates {
                ({128/38038}, 3.6353396)
                ({256/38038}, 4.09319052)
                ({512/38038}, 4.52715014)
                ({1024/38038}, 4.9655461)
                ({2048/38038}, 5.12293704)
                ({4096/38038}, 5.17871482)
            };
            \addlegendentry{Cross-PPI}

            \end{axis}
        \end{tikzpicture}
    }
    \caption{(a) Illustration of Cross-PPI \cite{sifaou2024semi}:
    The real data is divided into $K$ folds, and $K$ calibration procedures are conducted, each using all real data except one fold, held out for bias estimation. The calibration bias is estimated and removed from the synthetic data loss (as shown in Fig.~\ref{subfig:method_ppi}), obtaining the Cross-PPI loss.
    (b) Channel capacity as a function of the proportion of real data for Cross-PPI, PPI~\cite{angelopoulos2023prediction}, empirical risk minimization (ERM), and pseudo-empirical risk minimization (P-ERM). As detailed in~\cite{sifaou2024semi}, given real data combined with synthetic samples, a classification model is trained to map UE-location to the optimal downlink beam index. The performance is measured in terms of achievable downlink channel capacity.}
    \label{fig:cross_ppi}
\end{figure*}

\section{Modeling the Sim-to-Real Gap on the Training Objective} \label{sec:ppi_dt}

Bayesian learning can help mitigate the sim-to-real gap by accounting for uncertainty at the level of the environment model.
However, due to model misspecification and limited computational complexity, a residual sim-to-real gap is unavoidable in practice.
This gap affects the quality of the synthetic data generated by the \gls{dt}, which in turn degrades the accuracy of the training objective as an estimate of the true population loss.
To address this problem, this section explores the use of \gls{ppi}~\cite{angelopoulos2023prediction}, which offers a powerful statistical tool for debiasing the training objective, irrespective of any level of residual sim-to-real gap on the \gls{dt}.

\subsection{Semi-Supervised Learning via Prediction-Powered Inference}

\Gls{ppi} supports the definition of a semi-supervised training methodology that leverages both real and synthetic data. To elaborate, suppose we want to train an \gls{ai} model for a certain task using mainly synthetic data generated by the \gls{dt}. Assume also that we also have a limited set of real data for the same task, but  not enough to train a high-capacity model from scratch. In classical {transfer learning}, one might pre-train on synthetic data and then fine-tune on real data~\cite{simeone2022machine}. 

In contrast, \gls{ppi} uses the real data not to train the model directly, but to estimate the bias of a training loss estimate based on synthetic data, as illustrated by the images within the parenthesis in Fig.~\ref{subfig:method_ppi}. It then adjusts the model training objective (leftmost image in Fig.~\ref{subfig:method_ppi}) to remove this bias. Essentially, the real data provides a yardstick for how far off the synthetic training loss is from the real training loss.
\Gls{ppi} is appealing for telecommunication \gls{ai} problems because obtaining even a small validation set of real data is often feasible through field tests or pilot deployments, whereas gathering a comprehensive training set is not~\cite{sifaou2024semi}.

The \gls{ppi}-corrected training objective is unbiased with respect to the real data distribution and uses real data efficiently: instead of brute-force training, real samples are used to estimate simulator bias, while cheaper synthetic data provide most of the training signal. This enables good performance even when only a small fraction, e.g. $1$-$10\%$, of the training data is real~\cite{sifaou2024semi}. That said, \gls{ppi} in its basic form has two key limitations for telecommunication systems:

\noindent $\bullet$ \textbf{Data efficiency:} A subset of real data must be reserved solely for bias correction and cannot be used to calibrate the \gls{dt}, which is problematic when real data are extremely scarce.

\noindent $\bullet$ \textbf{Context independence:} Bias is typically corrected by a single global term, implicitly assuming the simulator error is the same in all contexts. In practice, \gls{dt} accuracy may vary by scenario, e.g., urban microcells vs. rural macrocells or across different signal-to-noise ratio ranges~\cite{ruah2025context}, so a single bias term cannot adequately correct all contexts.

\subsection{Cross-Validation-Based Prediction-Powered Inference}

Cross-PPI improves data efficiency by effectively recycling real data during calibration while still using it for bias estimation. As depicted in Fig.~\ref{fig:cross_ppi}(a), this is achieved through a cross-validation-style procedure in which the real data is partitioned into $K$ folds, and synthetic data is calibrated using $K$ calibration procedures, where each procedure uses all real data except one fold, reserved for bias correction~\cite{sifaou2024semi}. In such a way, all real data is used for both calibration and bias correction.

To illustrate the advantages of Cross-PPI, we consider the problem of beam alignment via \glspl{ckm} in mmWave massive MIMO systems~\cite{zeng2021toward}. A \gls{ckm} is a site‑specific database of channel characteristics associated with the locations of both the transmitter and the receiver~\cite{zeng2021toward}. Recently, this concept has been leveraged to mitigate training overhead, offering a promising new approach for channel estimation and beam alignment~\cite{zeng2021toward}. 

The objective is to train a classification model that maps the UE location to the optimal downlink beam selected from a predefined finite codebook. Fig.~\ref{fig:cross_ppi}(b) shows the performance of the trained model in terms of achievable downlink channel capacity as a function of the fraction of real samples. Three baselines are considered: \gls{erm}, which ignores the synthetic data; \gls{p-erm}, which treats the synthetic data as ground‑truth samples; and \gls{ppi}~\cite{angelopoulos2023prediction}. The results indicate that Cross‑PPI achieves substantial improvements compared to all other methods. Notably, \gls{p-erm} does not yield any clear benefits over \gls{erm}, which relies exclusively on the real data. Although \gls{ppi} shows some improvements over the baselines, its performance remains lower than that of Cross‑PPI.

\begin{figure*}
    \centering
    \subfigure[Real data\hspace{-0.1cm}]{
        \includegraphics[trim={0.1in 0 1.4in 0}, clip, keepaspectratio=true, height=3.3cm]{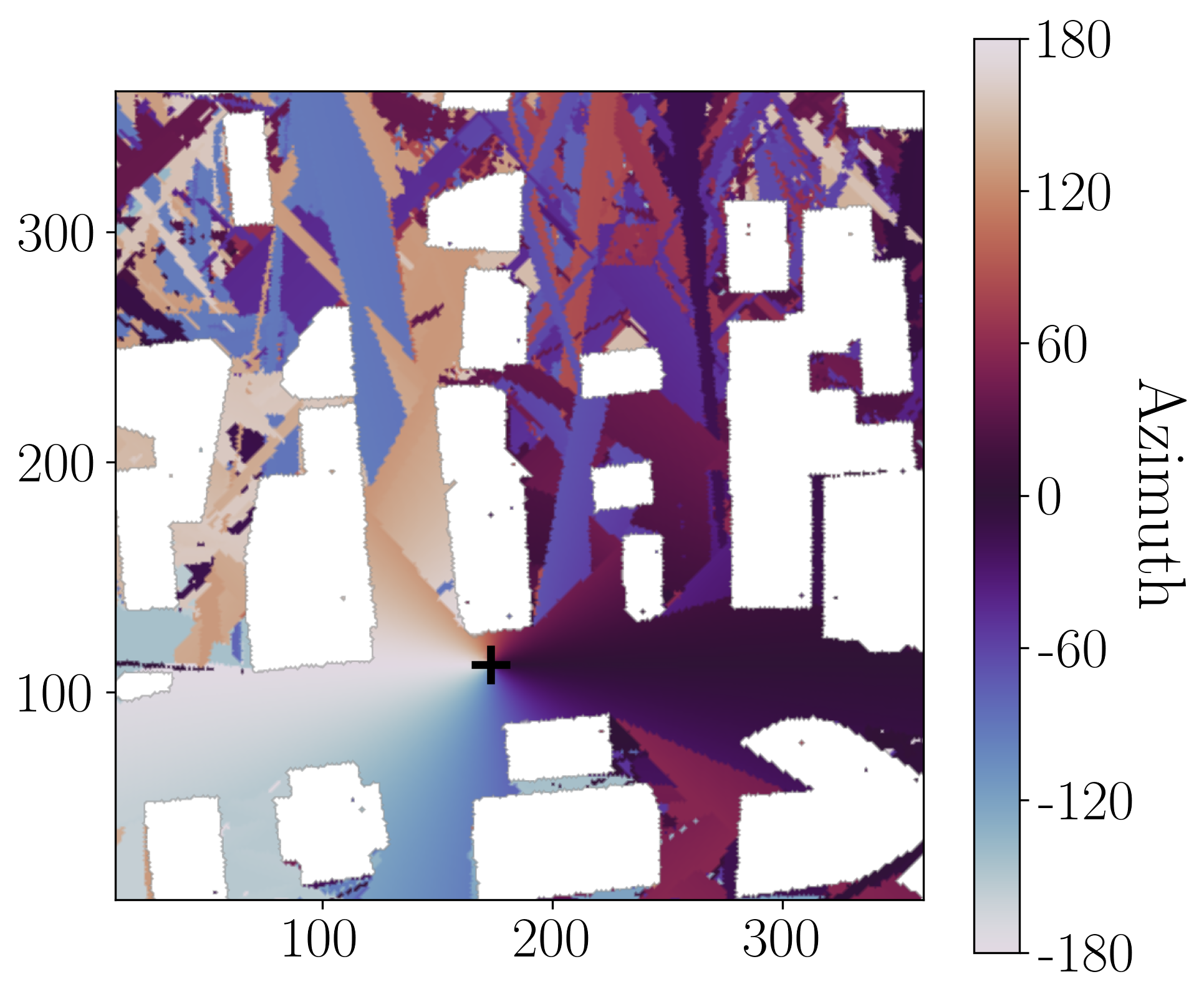}
        \label{subfig:gt_az}
    }
    \subfigure[Synthetic data\hspace{0.25cm}]{
        \includegraphics[trim={0.2in 0 0.4in 0}, keepaspectratio=true, height=3.3cm]{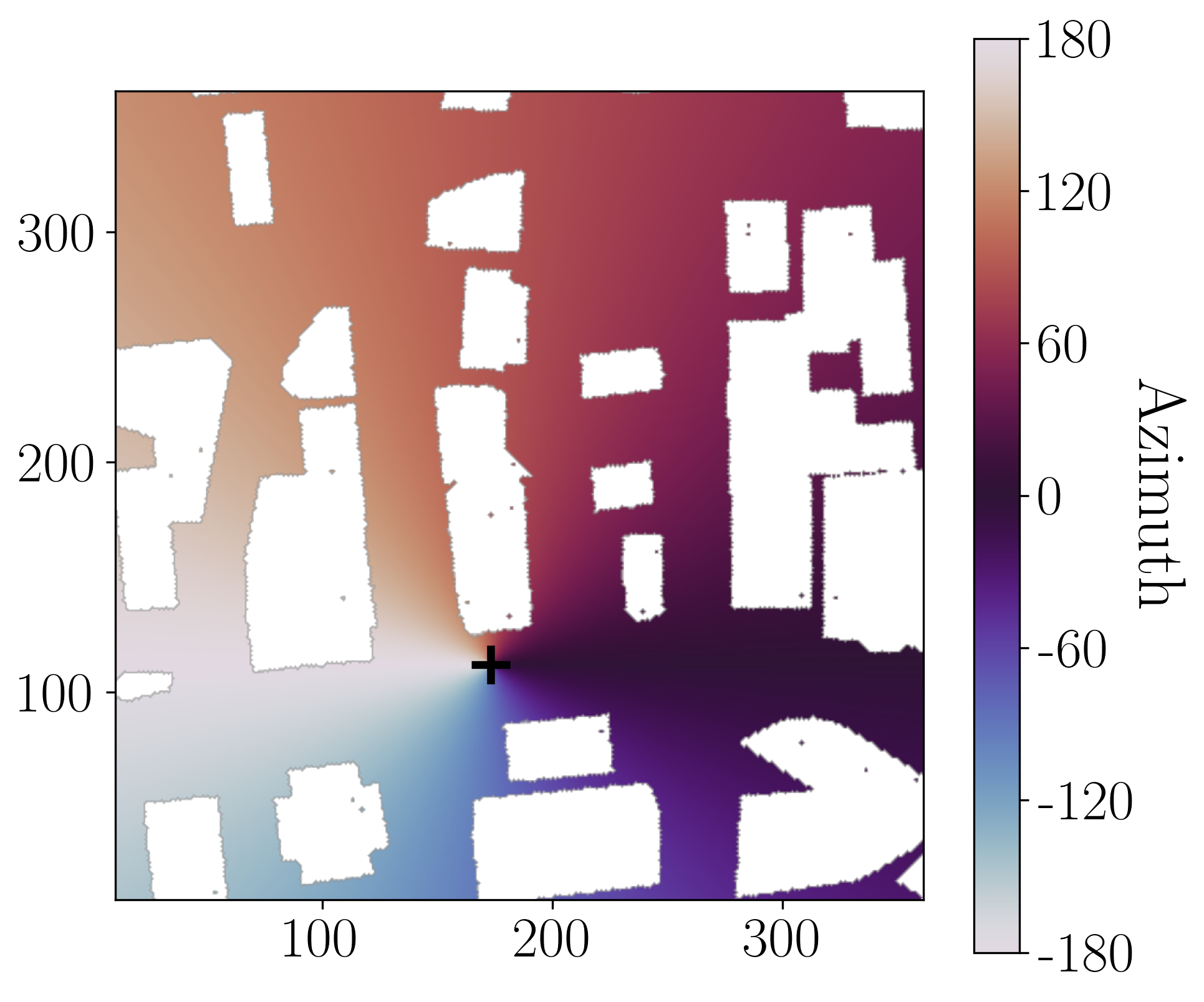}
        \label{subfig:teacher_az}
    }
    \subfigure[ERM\hspace{-0.3cm}]{
        \includegraphics[trim={0.1in 0 1.3in 0}, clip, keepaspectratio=true, height=3.15cm]{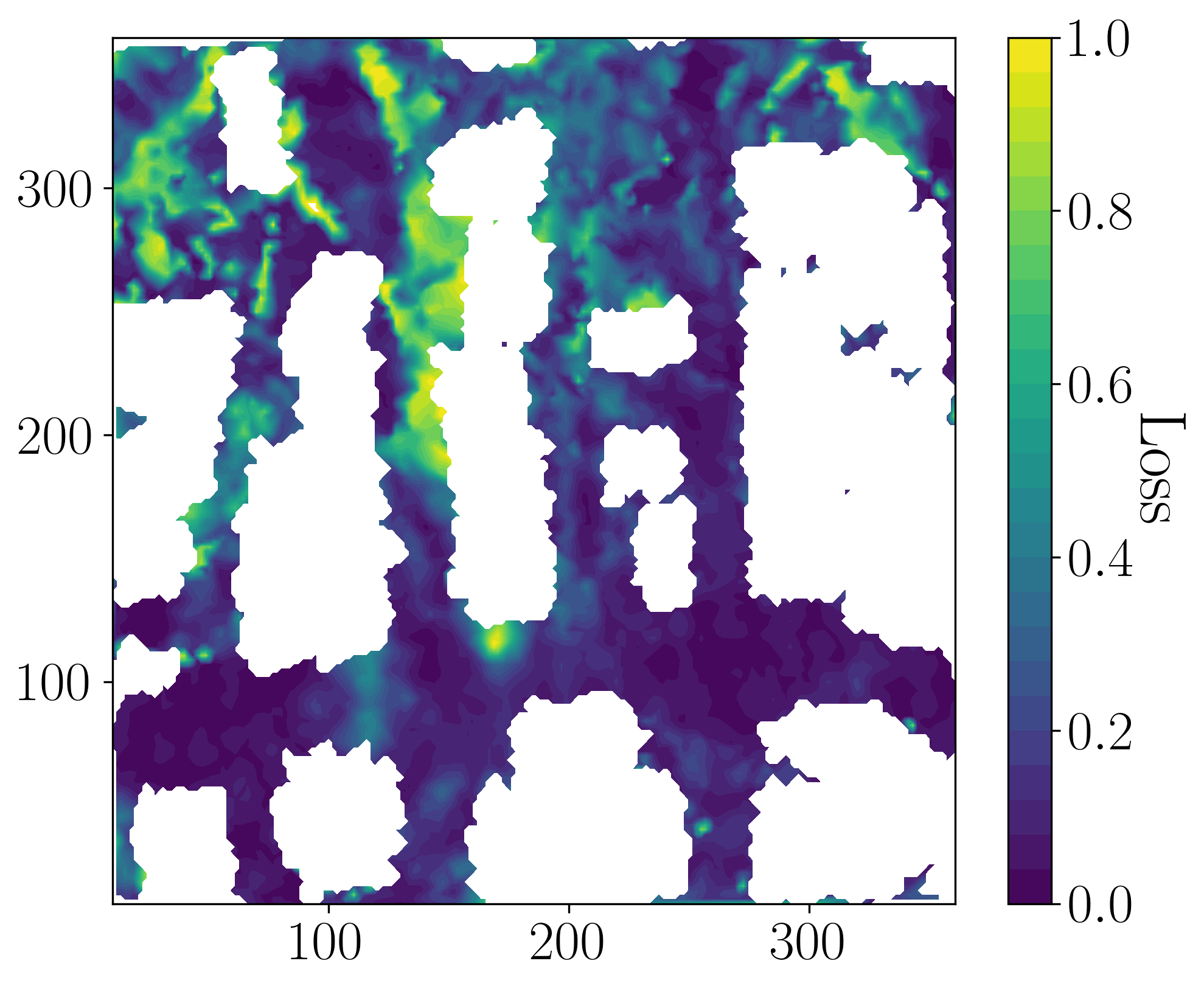}
        \label{subfig:beamfroming_loss_erm}
    }
    \subfigure[PPI\hspace{-0.25cm}]{
        \includegraphics[trim={0.1in 0 1.3in 0}, clip, keepaspectratio=true, height=3.15cm]{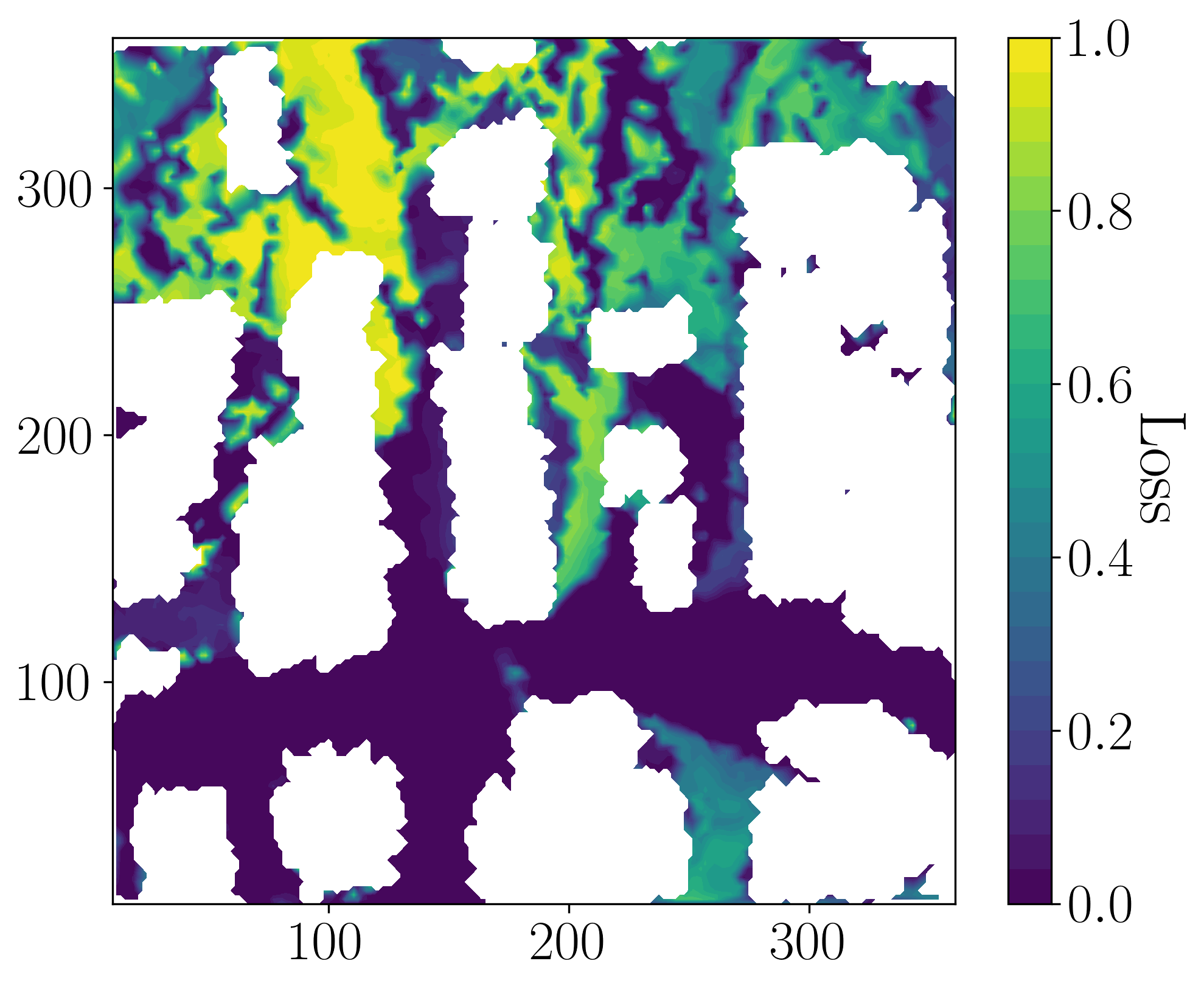}
        \label{subfig:beamfroming_loss_dr}
    }
    \subfigure[CPPI\hspace{0.1cm}]{
        \includegraphics[trim={0.2in 0 0.5in 0}, keepaspectratio=true, height=3.15cm]{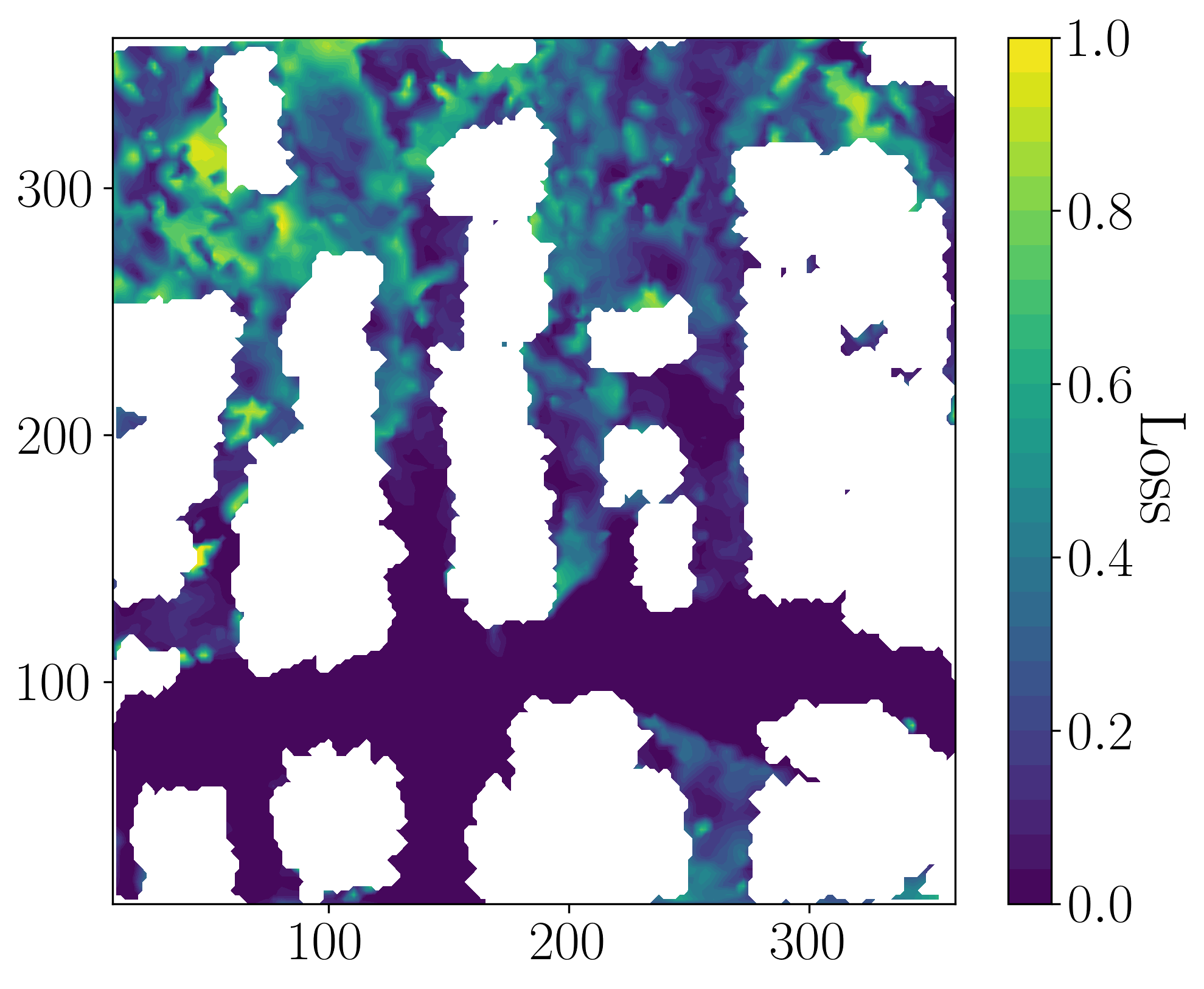}
        \label{subfig:beamfroming_loss_cdr}
    }
    \caption{(a) Heatmap for the ground-truth optimal azimuth angles required to serve a user in the given position from the BS placed at the black cross mark.  The layouts of the buildings are depicted as empty white regions. (b) Optimal azimuth angles as predicted by a DT. (c)-(e) Heatmap for the pointing error obtained by a  model trained using real and synthetic data via (c) empirical risk minimization (ERM), (d) PPI~\cite{angelopoulos2023prediction}, and (e) context-aware PPI (CPPI) \cite{ruah2025context}. }
    \label{fig:contextual_ppi}
\end{figure*}

\subsection{Context-Aware Prediction-Powered Inference}

Context-aware \gls{ppi}  tackles the lack of contextuality of conventional \gls{ppi} by allowing the bias correction applied to the training loss to vary with context. The core idea is to partition the synthetic data, and corresponding real validation data, as a function of relevant context features, and then perform bias estimation separately for each context. For instance, one can assign different correction factors for different ranges of a radio channel condition metric or for different cell sectors in a network. By doing so, the model treats synthetic data from contexts where the twin is reliable with more trust, and discounts or heavily corrects synthetic data in contexts where the twin is less accurate~\cite{ruah2025context}.

In Fig.~\ref{fig:contextual_ppi}, following \cite{ruah2025context}, we consider an outdoor urban environment with a single \gls{bs}, where the propagation paths between the \gls{bs} and the devices are generated using ray‑tracing as in~\cite{zeng2021toward}. Each device position is associated with a ground‑truth azimuth and elevation \gls{aod}, which is determined by selecting the propagation path with the highest received power. The objective is to train a mapping that outputs an estimate of the optimal \gls{aod} at each receive location, given a real dataset and a much larger synthetic dataset. 
Each data point is augmented with a binary context variable indicating if the device is within the \gls{los} of the BS.
  
Panels (a) and (b) present geographic heatmaps of beamforming directions: panel (a) shows the ground truth obtained from real data, while panel (b) displays the prediction made by a \gls{dt}. Notably, substantial discrepancies emerge in the \gls{nlos} region, where the \gls{dt} fails to capture the true beamforming pattern, yielding a large sim-to-real gap. Panels (c)–(e) quantify the beam-pointing error for three training schemes: \gls{erm}, which relies solely on real data, \gls{ppi}~\cite{angelopoulos2023prediction}, and \gls{cppi}~\cite{ruah2025context}, respectively. As seen, \gls{ppi} performs poorly in the \gls{nlos} locations, exhibiting even worse performance than \gls{erm}. In contrast, \gls{cppi} leverages context-dependent weightings of synthetic samples, thus achieving uniformly lower errors across the entire region.

\section{Conclusion} \label{sec:conclusion}

As \gls{ai} continues to scale, its promise for revolutionizing network automation and optimization is tempered by the realities of deployment in complex, data-scarce telecommunication environments. \Glspl{dt} have emerged as a vital tool to generate the data needed to train and test \gls{ai} solutions virtually. By calibrating \glspl{dt} with real-world data, and by explicitly modeling residual sim-to-real gaps through Bayesian and \gls{ppi} techniques, engineers can enhance the performance and reliability of \gls{dt}-aided telecommunication networks.
Practitioners may decide on different combinations of the techniques presented in this work depending on the type and quality of the available data and simulator (see Table~\ref{tab:methods_summary} for a summary).

We conclude by briefly discussing challenges and further topics.

\noindent \textbf{Real-time \glspl{dt}.} When \glspl{dt} are used for real-time tasks, a fundamental tension exists between \gls{dt} fidelity and computational efficiency: high-fidelity simulations provide accurate data but may be too slow for time-critical applications, while fast emulation models sacrifice accuracy \cite{hashash2022edge, pegurri2024toward}.

\noindent \textbf{Cross-site transferability and domain adaptation.} Calibrated \glspl{dt} are inherently site-specific, and the application of a calibrated \gls{dt} from one location or configuration to another typically requires the use of transfer learning and domain adaptation techniques.

\noindent \textbf{Continuous adaptation.} In practice, the design and calibration of a \gls{dt} for data augmentation must be updated either periodically or in response to changes in the environment, benefiting from frameworks such as continual learning and meta-learning \cite{simeone2022machine, chen2023learning}.

\bibliographystyle{IEEEtran}
\bibliography{biblio}

\vspace{-3em}
\begin{IEEEbiographynophoto}%
{Clement Ruah} is a Research Associate at the Institute for Intelligent Networked Systems at Northeastern University London.
He received his Ph.D. degree in Information Engineering from King's College London in 2025.
His research focuses on reliable machine learning and mixing simulation-based approaches with uncertainty-aware data-driven solutions.\\

\noindent\textbf{Houssem Sifaou} (Member, IEEE) received the M.S. and Ph.D. degrees in Electrical Engineering from the King Abdullah University of Science and Technology (KAUST) in 2016 and 2021, respectively.
He is currently a Research Fellow at the Institute for Intelligent Networked Systems at Northeastern University London. His research lies at the intersection of machine learning and wireless communications.\\

\noindent\textbf{Osvaldo Simeone} (Fellow, IEEE) is a Professor of Information Engineering at Northeastern University London, where he co-directs the Institute for Intelligent Networked Systems, and he is also a Visiting Professor at Aalborg University.
He has received multiple best paper awards, and his research has been supported by the NSF, ERC, EPSRC, European Commission, ESA, and industry.
He held previous positions at King's College London and at the New Jersey Institute of Technology.
Prof. Simeone is a Fellow of the IEEE, IET, and EPSRC, and the author of two textbooks published by Cambridge University Press.\\

\noindent\textbf{Bashir M. Al-Hashimi} worked in the electronics design industry for eight years prior to embarking on an academic career with the School of Electronics and Computer Science, University of Southampton, in 1999.
He is currently Vice President (Research \& Innovation) at King’s College London and co-Director of the King's Institute for Artificial Intelligence.
He has successfully led a number of large-scale interdisciplinary research programmers funded by the EPSRC and industry, supervised 40 Ph.D. students to successful completion, published 380 refereed technical papers, and authored or coauthored seven books.
\end{IEEEbiographynophoto}

\end{document}